\documentclass[aps,pra,reprint,amsmath,amssymb,superscriptaddress,nobibnotes, longbibliography]{revtex4-1}
\usepackage{graphicx,SIunits}
\usepackage{bm}     
\usepackage{hyperref} 
\usepackage{dsfont}    
\usepackage{color}

\usepackage{amsmath}
\usepackage{amsthm}
\usepackage{amssymb}
\usepackage{tikz}

\usetikzlibrary{decorations.pathreplacing}
\usepackage{braket}
\usepackage{dsfont}

\usepackage{color}

\begin{document}
	
	\title{Implementation of CU gates and its application in a remote-controlled quantum operation}
 
	\author{Byungjoo Kim}
	\email{These authors contributed equally}
	\affiliation{Center for Quantum Information, Korea Institute of Science and Technology (KIST), Seoul, 02792, Korea}
	\affiliation{Institute of Physics and Applied Physics, Yonsei University, Seoul 03722, Korea}
	\affiliation{Department of Laser \& Electron Beam Technologies, Korea Institute of Machinery and Materials (KIMM), Daejeon, 34103, Korea}
		
	\author{Seongjin Hong}
	\email{These authors contributed equally}
	\affiliation{Department of Physics, Chung-Ang University, Seoul 06974, Korea}
	
	\author{Yong-Su Kim}
	\affiliation{Center for Quantum Information, Korea Institute of Science and Technology (KIST), Seoul, 02792, Korea}
	\affiliation{Division of Quantum Information, KIST School, University of Science and Technology, Seoul, 02792, Korea}

        \author{Kyunghwan Oh}
	\affiliation{Institute of Physics and Applied Physics, Yonsei University, Seoul 03722, Korea}
	
	\author{Hyang-Tag Lim}
	\email{hyangtag.lim@kist.re.kr}
	\affiliation{Center for Quantum Information, Korea Institute of Science and Technology (KIST), Seoul, 02792, Korea}
	\affiliation{Division of Quantum Information, KIST School, University of Science and Technology, Seoul, 02792, Korea}
	
\date{\today} 
	
\begin{abstract}
Recently, remote-controlled quantum information processing has been proposed for its applications in secure quantum processing protocols and distributed quantum networks. For remote-controlled quantum gates, the experimental realization of controlled unitary (CU) gates between any quantum gates is an essential task. Here, we propose and experimentally demonstrate a scheme for implementing CU gates between arbitrary pairs of unitary gates using the polarization and time-bin degrees of freedom of single-photons. Then, we experimentally implement remote-controlled single-qubit unitary gates by controlling either the state preparation or measurement of the control qubit with high process fidelities. We believe that the proposed remote-controlled quantum gate model can pave the way for secure and efficient quantum information processing.
\end{abstract}
	
\maketitle

\section{Introduction}	
Quantum entanglement is a key concept in quantum physics and plays an essential role in the field of quantum information processing~\cite{Einstein1935, Kimble2008, Horodeck2009, Giovannetti2011}. Among its various applications, entangled operations, such as the controlled unitary (CU) gate including the controlled-NOT gate, are of particular significance, serving as a cornerstone for universal quantum computing~\cite{Nielsen2000, Knill2011, Barz2015}. Due to its importance, numerous efforts have been devoted to implement CU gates in different physical platforms including trapped ion~\cite{Schmidt2003, Leibfried2003}, superconducting~\cite{Steffen2006, Plantenberg2007}, atomic~\cite{Mandel2007, Anderlini2007}, and photonic systems~\cite{Obrien2003, Gasparoni2004, Bao2007}. One distinguishing feature of CU gates in photonic systems is their intrinsic ability for gate teleportation, enabling the realization of nonlocal quantum gates~\cite{Wan2019,Chou2018,Huang2004}. Hence, the implementation of CU gates in a linear optical system is essential for quantum networks, including distributed quantum computing and the quantum internet~\cite{Kimble2008,Simon2017,Humphreys2018}.

Recently, remote-controlled quantum computing (RCQC) has been proposed due to its potential for secure quantum processing, blind quantum computing~\cite{Barz2012}, distributed quantum network~\cite{Humphreys2018}, and distributed quantum sensing~\cite{Kim2024,Liu2021}. In RCQC, the client has a control qubit, while the server has a target qubit. The tasks of the client are remotely conducted as linear combinations of quantum gates performed by the server. These tasks are encoded as linear coefficients of arbitrary linear combinations of quantum gates. These linear coefficients are decomposed and encoded into a control qubit, enabling remote control of these operations by utilizing CU gates. The proof-of-principle experiments of RCQC were demonstrated by controlling the linear combination of two single-qubit gates using linear optics in the polarization basis~\cite{Wang2020, Qiang2017}. In addition, remote control of both the quantum state and operations on the target qubit was realized through the implementation of controlled gates using multiple degrees of freedom of sinlge-photons~\cite{Wang2020}. In previous studies, the quantum gates acting on the target qubit are remotely controlled by only manipulating the quantum state of control qubit. Moreover, in most cases, controlled gates are implemented by using maximally entangled states (Bell states) as input states or ancillary states~\cite{Obrien2003,Gasparoni2004,Pittman2003}. Recently, it has been reported that arbitrary entangled operations can be realized based on a coherent superposition of local operations in unbalanced interferometers, and arbitrary entangled operations can be acted on any quantum states including separable two-qubit states~\cite{Hong2020}. 


In this work, we experimentally demonstrate two-qubit controlled unitary gates and remote-controlled quantum gates on a single-qubit using the polarization and time-bin degrees of freedom of single-photons using unbalanced interferometers. We implement controlled-NOT (CNOT) and controlled-S (CS) gates with high process fidelities and then experimentally demonstrate remote-controlled single-qubit gates that realize a linear combination of Pauli operators. The linear coefficients of the two single-qubit unitary gates ($\alpha$ and $\beta$) acting on the target qubit are encoded either in the quantum state or in the projection operator of the control qubit. These linear coefficients of unitary gates can then be remotely controlled with high process fidelities through state preparation or measurement of the control qubit. We believe that our scheme provides a useful platform for secure quantum processing and distributed quantum networks.

\section{Theory}

\begin{figure}[h]
\centerline{\includegraphics[width=9cm]{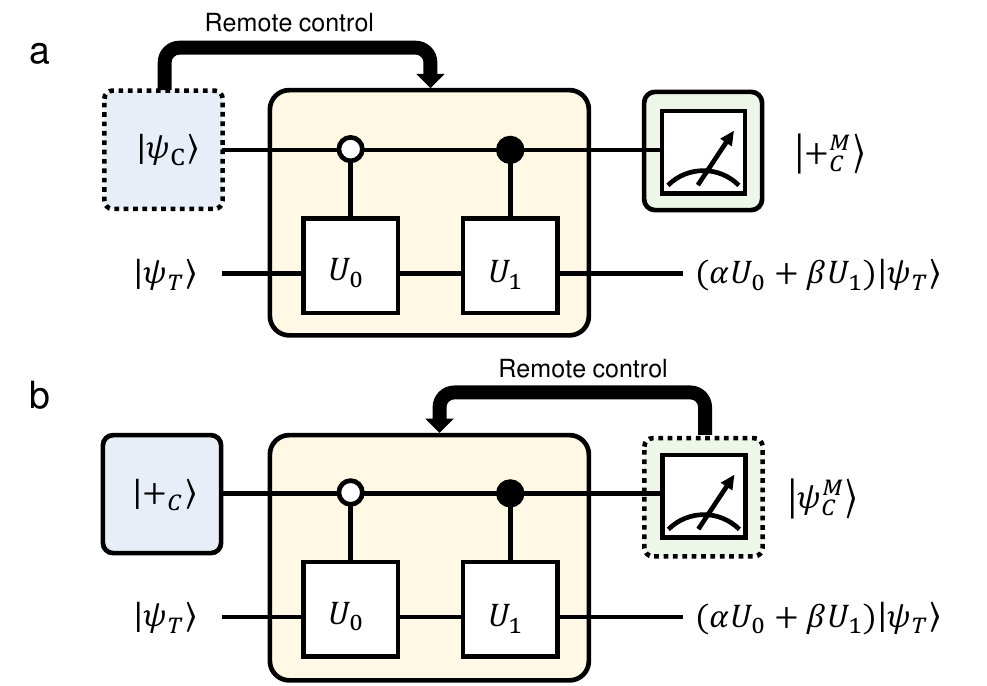}}
\caption{Quantum circuit diagram for a remote-controlled single-qubit operation $U_{RC}=\alpha U_0+\beta U_1$ by exploting controlled unitary gates. The linear coefficients $\alpha$ and $\beta$ can be remotely controlled by manipulating either (a) the quantum state or (b) the projection operator of the control qubit.}
\centering
\label{fig:figure1}
\end{figure}

A CU gate is a two-qubit gate that operates on two-qubit in a way that the first and second qubits serve as a control and a target, respectively~\cite{Nielsen2000}. A CU gate can be represented in the matrix form $|0\rangle\langle 0|\otimes U_0 + |1\rangle\langle 1|\otimes U_1 $, which acts on two-qubit states $|\psi_{{\rm C}}\rangle|\psi_{{\rm T}}\rangle$ where  $|\psi_{{\rm C}}\rangle$ and $|\psi_{{\rm T}}\rangle$ correspond to a control qubit and a target qubit, respectively. Here, $U_0$ and $U_1$ denote the arbitrary single-qubit unitary gates. The CU gate allows the control qubit $|\psi_{{\rm C}}\rangle$ to compute a unitary operation either $U_0$ or $U_1$, which operates on the target qubit $|\psi_{{\rm T}}\rangle$. In detail, if the state of the control qubit is $|0\rangle$ ($|1\rangle$), unitary operator $U_0$ ($U_1$) is applied to the target qubit $|\psi_{{\rm T}}\rangle$. 

Then we consider the concept of a remote-controlled quantum gate using the CU gates. In the quantum circuit model, the control qubit enables remote control of the application of an operation $U_{RC}$ on the target qubit. The $U_{RC}$ can be represented in a linear combination of two single-qubit unitary gates, i.e. $U_{RC}=\alpha U_{0}+\beta U_{1}$, where $\alpha$ and $\beta$ are complex numbers satisfying $|\alpha|^2+|\beta|^2=1$. The linear coefficients $\alpha$ and $\beta$ can be then encoded in a quantum state or projection operator of the control qubit. Here, we proposed two quantum circuit models of remote-controlled single-qubit operation by manipulating either the quantum state or projection operator of the control qubit as shown in Fig.~\ref{fig:figure1}.

At first, we consider a remote-controlled quantum gate by manipulating the quantum state of the control qubit as shown in Fig.~\ref{fig:figure1}(a). The initial two-qubit state is given as $|\Psi_{CT}\rangle=|\psi_C\rangle|\psi_T\rangle$ with $|\psi_C\rangle=\alpha |0\rangle+\beta |1\rangle$. After performing a CU gate, the two-qubit state is transformed to $|\Psi_{CT}\rangle=\alpha|0_C\rangle U_0|\psi_T\rangle+\beta |1_C\rangle U_1|\psi_T\rangle$. When the control qubit is projected onto $|+^M_C\rangle=(|0_C\rangle+|1_C\rangle)\sqrt{2}$, the target qubit state collapses onto

\begin{equation}
	\langle +^M_C|\Psi_{CT}\rangle=\frac{1}{\sqrt{2}}(\alpha U_0|\psi_T\rangle+\beta U_1|\psi_T\rangle).
\label{eq1}
\end{equation}
Thus, a single qubit unitary gate $U_{RC}=\alpha U_0+\beta U_1$ is acted on the target qubit $|\psi_T\rangle$. Note that $\alpha$ and $\beta$ can be remotely-controlled by adjusting the quantum state of the control qubit $|\psi_C\rangle$.

On the other hand, we now consider another quantum circuit model of a remote-controlled quantum gate by adjusting the projection operator of the control qubit as shown in Fig.~\ref{fig:figure1}(b). The initial state of control qubit is fixed as $|+_C\rangle$, and a CU gate is performed on the initial two-qubit state $|\Psi_{CT}\rangle=|+_C\rangle|\psi_T\rangle$, then $|\Psi_{CT}\rangle$ is transformed to $(|0_C\rangle U_0|\psi_T\rangle+|1_C\rangle U_1|\psi_T\rangle)\sqrt{2}$. Then, the control qubit is projected onto the measurement basis $|\psi^M_C\rangle=\alpha |0_C\rangle+\beta |1_C\rangle$ and the final state of the target qubit is collapses onto 

\begin{equation}
	\langle \psi^M_C|\Psi_{CT}\rangle=\frac{1}{\sqrt{2}}(\alpha U_0|\psi_T\rangle+\beta U_1|\psi_T\rangle).
	\label{eq2}
\end{equation}

One can find that the same form of the remote-controlled quantum gate $U_{RC}=\alpha U_0+\beta U_1$ is applied on target qubit $|\psi_T\rangle$, see Eq. (\ref{eq1}). In this quantum circuit, the linear coefficients of $\alpha$ and $\beta$ also can be remotely-controlled by adjusting the projection operator of the control qubit when the initial state of the control qubit is fixed with $|+_C\rangle$.

\begin{figure*}[t]
\includegraphics[width=5.5 in]{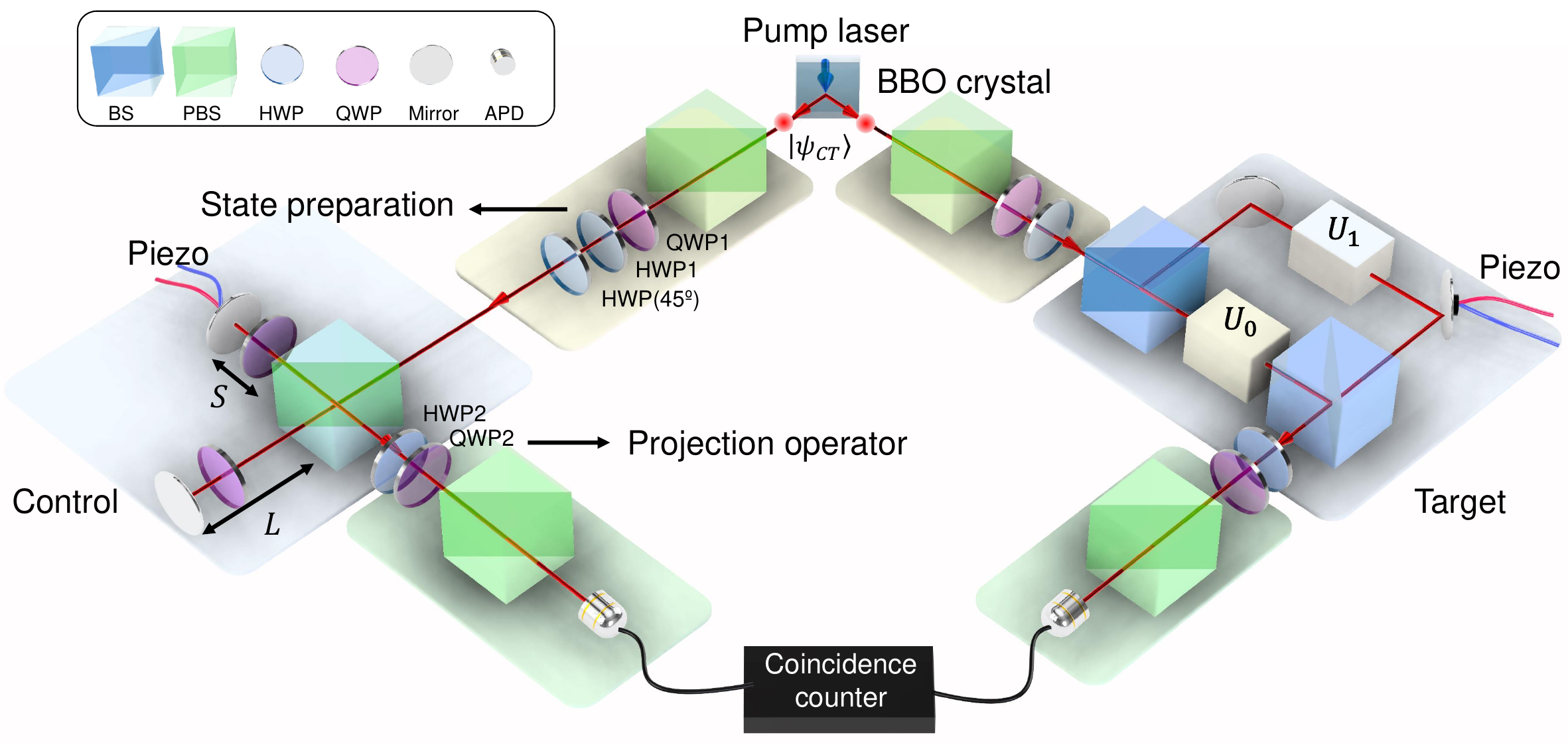} 
\caption{Experimental setup for implementing a remote-controlled quantum operation using a CU gate. A CU gate acting on a two-qubit polarization state of a down-converted photon pair is realized using an unbalanced Michelson interferometer (control) and an unbalanced Mach-Zehnder interferometer (target). With the implemented CU gate, a remote-controlled quantum operation on the target qubit can be demonstrated by adjusting either the initial state or the projection operator of the control qubit.}
\centering
\label{fig:figure2}
\end{figure*}

\section{Experiment}

We implement a CU gate using polarization and time-bin degrees of freedom of sinlge-photons~\cite{Hong2020}. Our proposed experimental setup is shown in Fig.~\ref{fig:figure2}, which is composed of two unbalanced interferometers. We use a mode-locked Ti-sapphire laser with a center wavelength of 780 nm, a repetition rate of 80 MHz, and a pulse width of 140 fs as a pump laser to prepare a down-converted photon pair. The initial pump beam is converted into 390 nm at a 1-mm-thick lithium triborate (LBO) crystal via a second harmonic generation process, and a pair of photons are generated from a 1-mm-thick beta-barium borate (BBO) crystal with a degenerate type-II spontaneous parametric down conversion (SPDC) process. Then, the signal photon is incident on the unbalanced Michelson interferometer (UMI), which is composed of a polarizing beam splitter (PBS), while the idler photon is incident on the unbalanced Mach-Zehnder interferometer (UMZI), which is composed of two beam splitters (BSs). Note that the signal (idler) photon refers to the control (target) qubit in our experiment. To actively stabilize the relative phase of the unbalanced interferometer, we attach a piezo actuator to one mirror of each interferometer. The path length difference of the UMI is set to 1.875 m ($\Delta L=L-S$), which is half of the pulse period of the mode-locked pump laser while the path length difference of the UMZI is set to 3.75 m ($2 \Delta L $). Thus, the photon pairs generated from consecutive pump laser pulses are temporally overlapped. We use an interference filter of 3 nm full-width at half-maximum (FWHM) bandwidth, and the expected coherence length of our down-converted photon is $ 200~\mu$m.

Let us describe how we implement a CU gate $|0\rangle\langle 0|\otimes U_0 +|1\rangle\langle 1|\otimes U_1$. The operations $|0\rangle\langle 0|$ and $|1\rangle\langle 1|$ on the control qubit can be performed by passing the PBS in the UMI, while $U_0$ and $U_1$ on the target qubit can be realized by placing a combination of waveplates (WPs) at the short and long path arms of the UMZI as shown in Fig.~\ref{fig:figure2}, respectively. The initial polarization-encoded two-qubit state $|\Psi_{CT}\rangle$ can be described as $|\Psi_{CT}\rangle|t_C^0t_T^0\rangle$ considering its time-bin mode, where $|t_C^0 \rangle$ ($|t_T^0  \rangle$) denotes the initial time-bin mode of the control (target) qubit. 


The process of the proposed CU gate can be described as the following transformation:
\begin{widetext}
\begin{eqnarray}
|\Psi_{CT}\rangle|t_C^0t_T^0\rangle
	&\xrightarrow{{\rm UMI, UMZI}} &
	\frac{1}{2} [ (|0\rangle\langle 0| \otimes U_0) |\Psi_{CT}\rangle|t_C^s t_T^s\rangle + (|0\rangle\langle 0| \otimes U_1) |\Psi_{CT}\rangle |t_C^s t_T^l\rangle \nonumber\\
	& & + (|1\rangle\langle 1| \otimes U_0) |\Psi_{CT}\rangle|t_C^l t_T^s\rangle + (|1\rangle\langle 1| \otimes U_1) |\Psi_{CT}\rangle|t_C^l t_T^l\rangle ] \nonumber\\
	&\xrightarrow{\text{Post-selection}} &
	\frac{1}{2}(|0\rangle\langle 0| \otimes U_0 + |1\rangle\langle 1| \otimes U_1)  |\Psi_{CT}\rangle
	\label{eq:transform},
\end{eqnarray}
\end{widetext}
where $|t_T^l  \rangle$ ($|t_C^s \rangle$) denotes the time-bin mode of the target (control) qubit passing through the long (short) path. The first line of Eq.~(\ref{eq:transform}) is obtained after the initial two-qubit states passing through the first BS and PBS in unbalanced interferometers. Then, the second line of Eq.~(\ref{eq:transform}) is obtained after $|t^s_C t^s_T\rangle$ and $|t^l_C t^l_T\rangle$ terms are post-selected with a time-bin measurement operator as $|t^s_C t^s_T\rangle\langle t^s_Ct^s_T|+|t^l_Ct^l_T\rangle\langle t^l_Ct^l_T|$ by set the coincidence window smaller than the time difference between long and short path in unbalanced interferometers~\cite{Hong2020}. Finally, we can obtain the CU gate $|0\rangle\langle 0|\otimes U_0+|1\rangle\langle 1|\otimes U_1$ as shown in Eq.~(\ref{eq:transform}). We apply this proposed arbitrary CU gate to the remote-controlled quantum gate by controlling either the state preparation or projection operator of the control qubit.

\section{Results}
\subsection{Controlled-unitary gates}
\begin{figure*}[t]
\includegraphics[width=5.5 in]{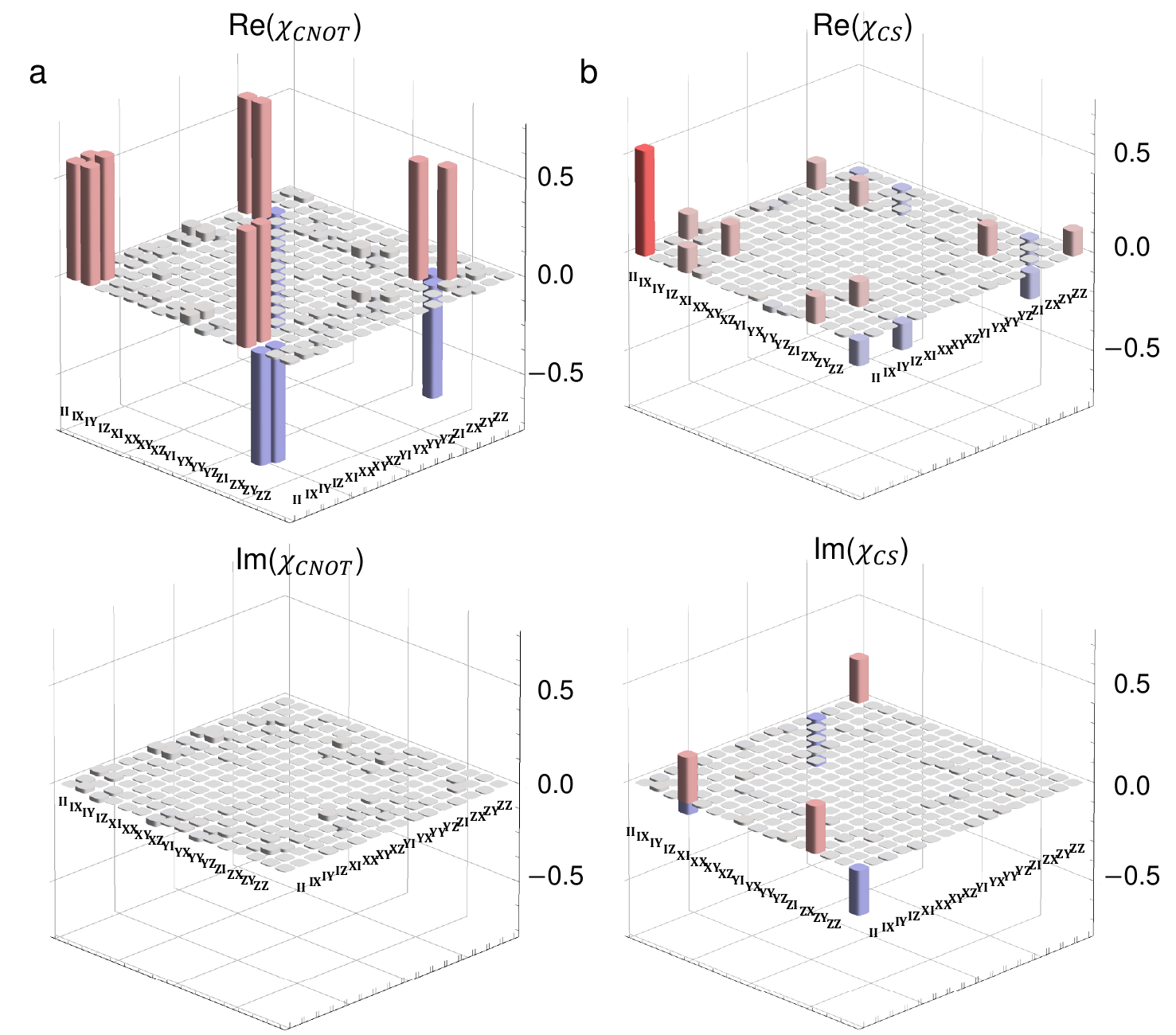}
\caption{Reconstructed process matrices of the experimentally reconstructed (a) CNOT gate and (b) CS gate. Re$(\chi)$ and Im$(\chi)$ correspond to the real and imaginary parts of the reconstructed process matrix $\chi$, respectively. Note that the reconstructed process matrices are represented on Pauli basis, for example, $I$, $X$, $Y$, and $Z$ correspond to $\sigma_{{I}}$, $\sigma_{{X}}$, $\sigma_{{Y}}$, and $\sigma_{{Z}}$, respectively.}
\centering
\label{fig:figure3}
\end{figure*}
We demonstrated CNOT and CS gates with our proposed CU gates. The CNOT gate can be realized by setting $U_0=I$, $U_1=X$. $I$ and $X$ operations are realized by placing nothing and a half waveplate (HWP) with an angle of $45^\circ$ at the short and long paths of the UMZI, respectively.  
In order to evaluate the qualities of implemented operations, we perform quantum process tomography (QPT). Figure 3(a) shows the experimentally reconstructed process matrix $\chi_{CNOT}$. The process fidelity of the experimental process matrix is obtained as 0.958$\pm$0.005. In order to evaluate the process fidelity of the realized operation, we use the formula $F(\chi_{{\rm exp}},\chi_{{\rm ideal}}) =\left[{\rm tr} (\sqrt{\sqrt{\chi_{{\rm exp}}} \chi_{{\rm ideal}}\sqrt{\chi_{{\rm exp}}}})\right]^2$ where $\chi_{{\rm ideal}}$ $(\chi_{{\rm exp}})$ denotes the process matrix of the ideal (experimentally realized) operation~\cite{Hong2020}. The standard deviation of the process fidelity is obtained by performing the Monte-Carlo simulation 100 times. In addition, the CS gate can be realized by setting $U_0=I$ and $U_1=S=\left( \begin{array}{cc} 1 & 0 \\ 0 & i \\ \end{array}  \right)$~\cite{ Schmidt2003, Garion2021}. $I$ and $S$ operations are implemented by placing nothing on the short path and QWP at $45^\circ$, HWP at $45^\circ$, and another QWP at $45^\circ$ on the long path in UMZI, respectively. Then, the relative phase was precisely adjusted with a piezo actuator to obtain coincidence counts corresponding to all state preparations and measurements. The reconstructed process matrix $\chi_{CS}$ is also shown in Fig. 3(b) with a process fidelity of 0.950$\pm$0.004. Note that the CS gate is a key two-qubit quantum gate used for implementing the quantum Fourier transform~\cite{Nielsen2000}. The CS gate is theoretically decomposed into the universal quantum gates of a $T$ gate, a $T^{-1}$ gate, and two CNOT gates, respectively~\cite{Mastriani2021}.

\subsection{Remote-controlled linear coefficients of unitary operators}

\begin{figure*}[t]
\includegraphics[width=5.5 in]{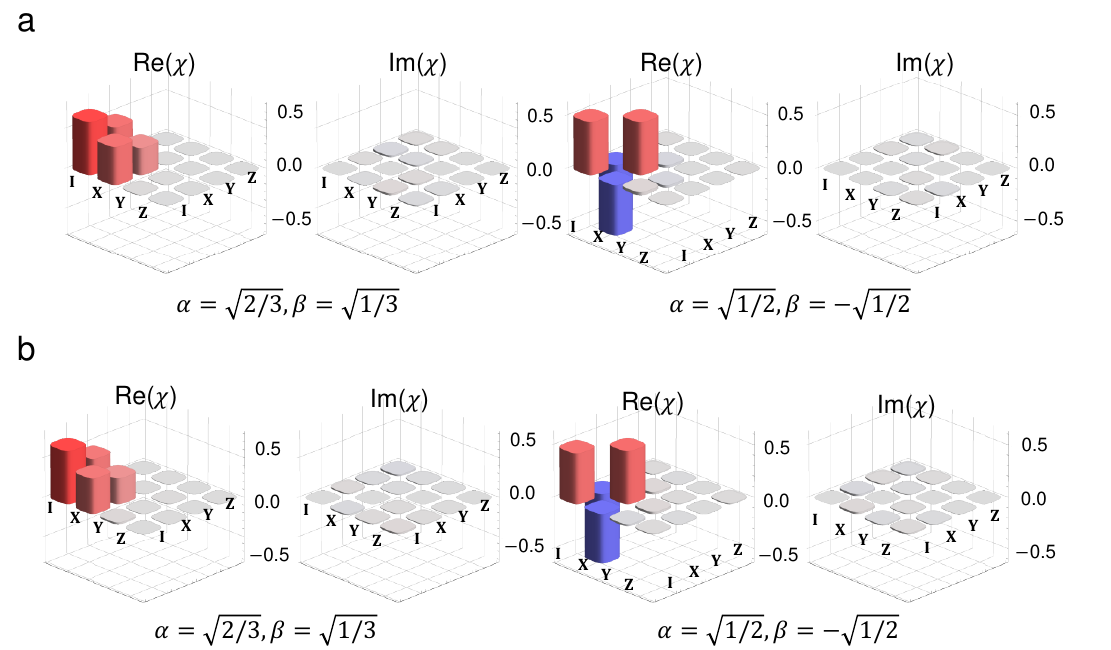}
\caption{Experimentally reconstructed process matrices of remote-controlled unitary operations $U_{RC}=\alpha I+\beta X$ acting on the target qubit. (a) Remote-controlled operation by manipulating the state preparation of the control qubit when the projection operator is fixed with $|+\rangle\ $. (b) Remote-controlled operation by manipulating the projection operator of the control qubit when the state preparation of the control qubit is fixed with $|+\rangle\ $. Re$(\chi)$ and Im$(\chi)$ refer to the real and imaginary parts of the reconstructed process matrix $\chi$, respectively.} 
\centering
\label{fig:figure4}
\end{figure*}

We demonstrate a remote-controlled quantum gate on the target qubit by controlling either the state preparation or projection operator of the control qubit by applying our proposed CU gates. First, we implement a remote-controlled quantum gate by manipulating the quantum state of the control qubit. The control qubit can be prepared as $|\psi_C\rangle=\alpha|0_C\rangle+\beta|1_C\rangle$. Note that the amplitudes $\alpha$ and $\beta$ are set by adjusting the angles of HWP1 and a quarter waveplate (QWP1), and the relative phase is set by adjusting the path length of the UMI with a piezo actuator embedded in the mirror as shown in Fig.~\ref{fig:figure2}. After applying the proposed CU gates as the final form of Eq.~(\ref{eq:transform}), the two-qubit quantum state becomes as $(\alpha  |0_C\rangle U_0 |\psi_T\rangle+\beta  |1_C\rangle U_1 |\psi_T\rangle)/\sqrt{2}$. When the control qubit is projected onto $|+_C^M\rangle=(|0\rangle+|1\rangle)/\sqrt{2}$  with HWP2, QWP2 and PBS after the UMI, then the quantum state collapses onto $(\alpha U_0+\beta U_1)|\psi_T\rangle$. Thus, a remote-controlled quantum gate, which is performed on the target qubit of $|\psi_T\rangle$, can be described as $U_{RC}=\alpha U_0+\beta U_1$. The linear coefficients of $\alpha$ and $\beta$ can be remotely controlled by preparing a quantum state of the control qubit $|\psi_C\rangle$. 

We experimentally demonstrate two remote-controlled quantum gates of $U^{S}_{RC,A}=\sqrt{2/3} I + \sqrt{1/3} X$ and $U^{S}_{RC,B}=\sqrt{1/2} I-\sqrt{1/2} X$ as examples. $I$ and $X$ single-qubit unitary operations are realized by placing nothing and HWP at $45^\circ$ in the short and long paths in the UMZI, respectively. To implement $U^{S}_{RC,A}$ ($U^{S}_{RC,B}$), the quantum states of controlled qubit were manipulated by setting the angles of HWP1 and QWP1 at $30^\circ$ and $0^\circ$ ($-22.5^\circ$ and $45^\circ$), respectively. The projection operator $|+_C^M\rangle \langle +_C^M |$ can be performed by setting the angles of HWP2 and QWP2 at $22.5^\circ$ and $45^\circ$, respectively. In order to estimate the quality of the implemented remote-controlled quantum gates, we perform QPT and the experimentally reconstruced process metrics are shown in Fig.~\ref{fig:figure4}(a). The corresponding process fidelities of the reconstructed process matrices are 0.998$\pm$0.002 and 0.974$\pm$0.005, respectively.


Then, we demonstrate a remote-controlled quantum gate by adjusting the projection operator of the control qubit when the quantum state of the control qubit is fixed as $|+_C\rangle=(|0\rangle+|1\rangle)/\sqrt{2}$. To this end, the control qubit is fixed as $|\psi_C\rangle=|+_C\rangle$, and the projection operator $|\psi^M_C\rangle \langle \psi^M_C |$ with $|\psi^M_C\rangle = \alpha |0 \rangle + \beta |1 \rangle $ is used for a control qubit. In this scheme, the same remote-controlled quantum operation $U_{RM}=\alpha U_0+\beta U_1$ can be performed on the target qubit, but the linear coefficients $\alpha$ and $\beta$ are remotely controlled by the projection operator of the control qubit, $|\psi^M_C\rangle \langle \psi^M_C |$. $U_0=I$ and $U_1=X$ unitary operations are realized in the same way as the above scheme. The quantum state of the control qubit $|\psi_C\rangle=|+_C\rangle$ is prepared by setting HWP1 at $22.5^\circ$ and QWP1 at $45^\circ$. Then, the projection operator of the control qubit $|\psi^M_C\rangle \langle \psi^M_C |$ can be prepared by adjusting the angles of HWP2 and QWP2. We demonstrate the same gates of $U^{P}_{RC,A}= \sqrt{2/3}I+ \sqrt{1/3}X$ and $U^{P}_{RC,B}= \sqrt{1/2}I-\sqrt{1/2}X$ by adjusting the angles of HWP2 and QWP2 when the control qubit is fixed as $|\psi_C\rangle=|+_C\rangle$. The angles of HWP2 and QWP2 are set as $30^\circ$ and $0^\circ$ ($-22.5^\circ$ and $45^\circ$) to implement $U^{P}_{RC,A}$ ($U^{P}_{RC,B}$), respectively. The experimentally reconstructed process matrices of $U^{P}_{RC,A}$ and $U^{P}_{RC,B}$ are shown in Fig.~\ref{fig:figure4}(b). The corresponding process fidelities of the reconstructed process matrices are 0.974$\pm$0.005 and 0.973$\pm$0.004, respectively. 

Our results show that by adjusting either the state preparation or projection operator of the control qubit, the linear coefficients of the remote-controlled quantum gates $U_{RC}$ applied to the target qubit can be remotely controlled as desired. In particular, it is noteworthy that the linear coefficients of unitary gates acting on a target qubit can be changed through measurements even after the quantum operation consistent with the theoretical prediction in Fig.~\ref{fig:figure1}(b). Note that the projection operator can be postponed even after the unitary operation has been finished. It is an intriguing direction to study the remote-controlled unitary operations with delayed-choice in the quantum circuits model~\cite{Peruzzo2012,Kaiser2012}.

\section{Conclusions}
We experimentally demonstrate two-qubit CU gates by exploiting polarization and time-bin degrees of freedom in a photonic system. Furthermore, we experimentally demonstrated a remote-controlled quantum gate model by controlling either the quantum state of the control qubit or the projection operator of the control qubit. Since our scheme can be extended to multi-qubit systems, it is intriguing to apply our proposed scheme to the controlled-controlled-unitary gate~\cite{Patel2016,Lanyon2009}. Moreover, we can consider the controlled non-unitary operation by replacing a single-qubit unitary operation with a projection operator or a generalized POVM~\cite{Lim2014, Hong2022, Choi2020}. Note that this controlled non-unitary operation can be used for the entanglement filter~\cite{Okamoto2009}. A remote-controlled quantum gate model can be directly applicable to the superposition of quantum gates~\cite{Procopio2015, Renner2022}, reducing query complexity~\cite{Araujo2014, Wei2016}, quantum network~\cite{Kimble2008, Wehner2018}, and Hamiltonian simulations~\cite{Childs2012, Berry2015}. This quantum circuit model can provide insights to secure remote quantum information processing, which is useful for clients concerned about information exposure during quantum information processing.


\end{document}